# A Reed Muller-based approach for optimization of general binary quantum multiplexers

Kevin Jin
Department of Electrical and Computer Engineering
Portland State University
Portland, USA
kevindujin@gmail.com

Tahsin Saffat
Department of Mathematics
Massachusetts Institute of Technology
Cambridge, USA
tahsin.saffat@gmail.com

Marek Perkowski
Department of Electrical and Computer Engineering
Portland State University
Portland, USA
h8mp@pdx.edu

*Abstract*—Previous work has provided methods for decomposing unitary matrices to series of quantum multiplexers, but the multiplexers created in this way are highly non-minimal. This paper presents a new approach for optimizing quantum multiplexers with arbitrary single-qubit quantum target functions. For quantum multiplexers, we define standard forms and two types of new forms: fixed polarity quantum forms (FPQF) and Kronecker quantum forms (KQF), which are analogous to Minterm Sum of Products forms, Fixed Polarity Reed-Muller (FPRM) forms, and Kronecker Reed-Muller (KRM) forms, respectively, for classical logic functions. Drawing inspiration from the usage of butterfly diagrams for FPRM and KRM forms, we devise a method to exhaustively construct all FPQF and KQF forms. Thus, the new forms can be used to optimize quantum circuits with arbitrary target unitary matrices, rather than only multi-controlled NOT gates such as CNOT, CCNOT, and their extensions. Experimental results on FPQF and KQF forms, as well as FPRM and KRM classical forms, applied to various target gates such as NOT, V, V+, Hadamard, and Pauli rotations, demonstrate that FPQF and KQF forms greatly reduce the gate cost of quantum multiplexers in both randomly generated data and FPRM benchmarks.

## I. INTRODUCTION

Previous works in the field of quantum compilation, such as [1], have generated methods for decomposition of arbitrary unitary matrices into a series of quantum multiplexers, but the multiplexers created by these methods are highly non-minimal. Also, some authors create parts of their oracles as logic functions represented in the EXOR of ANDs form, such as ESOPs, which are not necessarily minimal. Furthermore, descriptions of sequences of multi-controlled arbitrary unitary gates appear in several quantum algorithms, which may be non-optimal. Vartiainen et. al. [2] and other authors present methods that can generate arbitrary n-qubit gates, but the resultant circuits are very expensive. We discuss a special case of the method in [2], where the gates are controlled by many inputs, but target only one qubit. Our method is not as general as previous methods as it generates only standard n-qubit multiplexers, but for this special case, our results are less expensive, or even exactly minimal. This paper is intended to help improve previous work by providing more efficient methods of building multiplexers, enabling the previously developed multiplexer decomposition methods to become more practical.

## II. BACKGROUND

We first discuss Fixed Polarity Reed-Muller (FPRM) forms and Kronecker Reed-Muller (KRM) forms, their corresponding butterfly diagrams, and rudimentary cost calculations for Boolean functions in FPRM form realized through quantum circuits. By understanding FPRM and KRM butterflies, the butterfly diagrams that we create here for optimization of binary quantum multiplexers will become clear.

The concept of a quantum multiplexer was first introduced by Shende et. al. in [3], where they propose the quantum multiplexer circuit block for usage in recursive decompositions. Other work by Shende et. al. in [4] focused on optimization of two-qubit unitary operators, but not on optimization of larger circuits such as quantum multiplexers. In [2], Vartiainen et. al. demonstrate a method for optimization of arbitrary multi-qubit gates, but do not provide a method specifically for optimization of the less general multi-qubit multiplexers. Tucci developed the Qubiter program in [1] to decompose arbitrary unitary matrices to series of quantum multiplexers, but a later work by Hutsell [9] has shown that the output from Qubiter is highly non-minimal, leaving plenty of space for optimization.

### A. FPRM and Davio expansions, and KRM and Shannon expansions

FPRM forms are a canonical form of a Boolean function where all variables appear in at most one polarity (that is, their polarities are "fixed"). FPRM forms, as well as some other expressions, use the concept of polarity of variables. As an example, for some arbitrary three-variable binary Boolean function, let us choose the polarity 101; this represents the polarity of a function on variables $a, b$, and $c$ where $a$ appears uncomplemented, $b$ appears complemented, and $c$ appears uncomplemented. The FPRM for this polarity would have the form

$F = p_0 1 \oplus p_1 c \oplus p_2 \bar{b} \oplus p_3 \bar{b}c \oplus p_4 a \oplus p_5 ac \oplus p_6 a\bar{b} \oplus p_7 a\bar{b}c$

, where $p_i$ are the spectral coefficients (which are either 0 or 1), and the functions $1, c, \bar{b}, \bar{b}c, a$, etc. are called the base functions.

The FPRM form is closely related to the positive and negative Davio expansions (see [5] and [6]) of functions on single variables. To better understand Davio expansions, it is helpful to first introduce Shannon expansions. The Shannon expansion of a Boolean expression splits the expression into two disjoint parts (called cofactors), where each cofactor results from substituting either 0 or 1 for a particular variable in the original function. For example, for the expression $F = \bar{a} + ab$, a Shannon expansion on variable $a$ would result in the general form $F = \bar{a}F_{\bar{a}} + aF_a$, where $F_{\bar{a}}$ is the negative cofactor, which results from evaluating $F$ for $a = 0$ (in this case, $F_{\bar{a}} = 1 + 0b = 1$). Likewise, $F_a$ is the positive cofactor, which results from evaluating $F$ for $a = 1$ (in this case, $F_a = 0 + 1b = b$). By substituting these cofactors back into the Shannon expansion, we can verify that it is the same as the original function.

The positive and negative Davio expansions are closely related to the Shannon expansion, except that the Davio expansions require that the variable only appears in either uncomplemented form (positive Davio) or complemented form (negative Davio). First, replace the OR in the Shannon expansion with an EXOR since $\bar{a}F_{\bar{a}}$ and $aF_a$ are disjoint; this changes the expansion from $F = \bar{a}F_{\bar{a}} + aF_a$ to $F = \bar{a}F_{\bar{a}} \oplus aF_a$. Both of the Davio expansions can be directly derived from this EXOR form of the Shannon expansion by substituting $(1 \oplus a)$ for $\bar{a}$ (to generate a positive Davio expansion), or by substituting $(1 \oplus \bar{a})$ for $a$ (to generate a negative Davio expansion). From this substitution, we obtain the general positive Davio form $F = (1 \oplus a)F_{\bar{a}} \oplus aF_a$ and the general negative Davio form $F = \bar{a}F_{\bar{a}} \oplus (1 \oplus \bar{a})F_a$. It can be shown that the positive Davio form can be rearranged as $F = 1(F_{\bar{a}}) \oplus a(F_{\bar{a}} \oplus F_a)$, and the negative Davio form can be rearranged as $F = \bar{a}(F_{\bar{a}} \oplus F_a) \oplus 1(F_a)$. Note that a series of Davio expansions on each variable of a general Boolean expression allows us to generate an FPRM form (the proof of which is beyond the scope of this paper); we will now give an example of this process.

Let us begin with an example function: $F(a,b) = \bar{a} + ab$, which we wish to convert to the FPRM form with polarity 10. We first apply a positive Davio expansion on variable $a: F = 1(1) \oplus a(1 \oplus b) = 1 \oplus a \oplus ab$. We then apply a negative Davio expansion on variable $b: F = 1(1) \oplus \bar{b}(\bar{a} \oplus 1) = 1 \oplus a\bar{b}$. At this point, we have the FPRM form for $F$ with polarity 10. By comparing our FPRM form to the general FPRM form with polarity 10: $F = p_0 1 \oplus p_1 \bar{b} \oplus p_2 a \oplus p_3 a\bar{b}$, we see that the spectral coefficients are $[p_0, p_1, p_2, p_3] = [1, 0, 0, 1]$, which can be expressed more succinctly as 1001.

KRM forms are similar to FPRM forms. In FPRM, each variable can be in either positive or negative polarity; in KRM, each variable can be in either positive, negative, or mixed polarity. While positive or negative polarity on a variable can be obtained by applying positive or negative Davio expansions, respectively, a mixed polarity on a variable can be obtained by applying the Shannon expansion on that variable. For example, let us begin with the same function: $F(a,b) = \bar{a} + ab$, which we wish to convert to the KRM form with polarity 12 (we use the symbol '2' to denote a mixed polarity). We first apply a positive Davio expansion on variable $a: F = 1(1) \oplus a(1 \oplus b) = 1 \oplus a \oplus ab$. We then apply a Shannon expansion on variable $b: F = \bar{b}(1 \oplus a) \oplus b(1) = \bar{b} \oplus a\bar{b} \oplus b$. At this point, we have the KRM form for $F$ with polarity 12. Given that the general KRM form with polarity 12 is $F = p_0 \bar{b} \oplus p_1 b \oplus p_2 a\bar{b} \oplus p_3 ab$, this means that the spectral coefficients are $[p_0, p_1, p_2, p_3] = [1, 1, 1, 0]$.

As will be seen in the next section, it is more convenient to use the vector representation of the spectral coefficients when dealing with transformation matrices. Additionally, a more in-depth overview of Reed-Muller families can be found in [10].

*B. Use of Kronecker product to generate transformation matrices*

For Boolean functions, minterms are product terms that use all literals of the function; for example, if a binary function uses variables $a, b$, and $c$, valid minterms include product terms such as $\bar{a}b\bar{c}, ab\bar{c}$, etc. Any function can be fully defined by the minterms that it covers; for example, the function $F = a + b$ can be fully defined as an OR of the three minterms it covers: $F = a\bar{b} + \bar{a}b + ab$. Such a representation

$$\begin{bmatrix} a & b \\ c & d \end{bmatrix} \otimes \begin{bmatrix} x & y \\ z & w \end{bmatrix} = \begin{bmatrix} ax & ay & bx & by \\ az & aw & bz & bw \\ cx & cy & dx & dy \\ cz & cw & dz & dw \end{bmatrix}$$

Fig. 1. *Example of Kronecker product ($\otimes$) on two 2x2 matrices*

of a function is known as a (canonical) Minterm Sum of Products form. Because the paper will use the concept of the Kronecker product, Figure 1 illustrates this concept. The Kronecker product can be applied to any two matrices, each of arbitrary dimension. Using the Kronecker product repeatedly, a transformation matrix is created that generates the spectral coefficients of an FPRM or KRM form. When given the minterms of the function as a vector, one can select the transformation matrix of the desired polarity and calculate the vector of spectral coefficient by multiplying the vector of minterms by the matrix. The minterms of the function are passed into the transformation matrix as a vector, in which the values of minterms are listed in ascending natural order. To create the transformation matrix, we use multiple applications of the Kronecker product: for each positive polarity variable, compose the **polarity transformation** $\begin{bmatrix} 1 & 0 \\ 1 & 1 \end{bmatrix}$ to the Kronecker product, for each negative polarity variable, compose the polarity transformation $\begin{bmatrix} 1 & 1 \\ 0 & 1 \end{bmatrix}$, and for each mixed polarity variable, compose the polarity transformation

$\begin{bmatrix} 1 & 0 \\ 0 & 1 \end{bmatrix}$. We provide intuitive reasoning for why these three specific transformations are used by applying them to a vector of arbitrary cofactors: $\begin{bmatrix} F_{\bar{a}} \\ F_a \end{bmatrix}$. Note that when matrices are multiplied in the context of binary values, addition modulo-2 (which is equivalent to EXOR) is used instead of normal addition when combining products of entries. Thus, if we apply the previously introduced positive transformation matrix to the cofactor vector, we see that $\begin{bmatrix} 1 & 0 \\ 1 & 1 \end{bmatrix} \begin{bmatrix} F_{\bar{a}} \\ F_a \end{bmatrix} = \begin{bmatrix} F_{\bar{a}} \\ F_{\bar{a}} \oplus F_a \end{bmatrix}$. Noting that a positive Davio expansion can be written as $F = 1(F_{\bar{a}}) \oplus a(F_{\bar{a}} \oplus F_a)$, we notice the similarities between the output of the transformation matrix and the positive Davio expansion; simply applying $1$ to the first output $(F_{\bar{a}})$ and $a$ to the second output $(F_{\bar{a}} \oplus F_a)$ allows us to generate the positive Davio expansion. Through this same process, a similar observation can be made about the close relationship between the negative polarity transformation $\begin{bmatrix} 1 & 1 \\ 0 & 1 \end{bmatrix}$ and the negative Davio expansion $F = \bar{a}(F_{\bar{a}} \oplus F_a) \oplus 1(F_a)$. A similar close relationship can be noted between the mixed polarity transformation $\begin{bmatrix} 1 & 0 \\ 0 & 1 \end{bmatrix}$ and the Shannon expansion $F = \bar{a}F_{\bar{a}} \oplus aF_a$. Finally, since we have previously demonstrated that repeated Davio expansions allow us to generate an FPRM form, and that repeated Davio and Shannon expansions allow us to generate a KRM form, and since the Kronecker product of transformation matrices (for reasons that are beyond the scope of this paper; see [7]) can emulate repeated Davio/Shannon expansions, this means that the Kronecker product of multiple transformation matrices also allows us to generate FPRM and KRM forms.

We now provide an example of how FPRM and KRM forms can be generated using the above method on the function $F = a \oplus b$ (with minterm vector $\begin{bmatrix} m_0 \\ m_1 \\ m_2 \\ m_3 \end{bmatrix} = \begin{bmatrix} 0 \\ 1 \\ 1 \\ 0 \end{bmatrix}$) expressed in FPRM polarity 10. The transformation $T$ would be $\begin{bmatrix} 1 & 0 \\ 1 & 1 \end{bmatrix} \otimes \begin{bmatrix} 1 & 1 \\ 0 & 1 \end{bmatrix} = \begin{bmatrix} 1 & 1 & 0 & 0 \\ 0 & 1 & 0 & 0 \\ 1 & 1 & 1 & 1 \\ 0 & 1 & 0 & 1 \end{bmatrix}$, and the vector of spectral coefficients would be the result of the matrix $T$ multiplied with the vector of minterms: $\begin{bmatrix} 1 & 1 & 0 & 0 \\ 0 & 1 & 0 & 0 \\ 1 & 1 & 1 & 1 \\ 0 & 1 & 0 & 1 \end{bmatrix} \begin{bmatrix} 0 \\ 1 \\ 1 \\ 0 \end{bmatrix} = \begin{bmatrix} 1 \\ 1 \\ 0 \\ 1 \end{bmatrix}$. Alternatively, if we wished to express the function in KRM polarity 20, the transformation $T$ would be $\begin{bmatrix} 1 & 0 \\ 0 & 1 \end{bmatrix} \otimes \begin{bmatrix} 1 & 1 \\ 0 & 1 \end{bmatrix} = \begin{bmatrix} 1 & 1 & 0 & 0 \\ 0 & 1 & 0 & 0 \\ 0 & 0 & 1 & 1 \\ 0 & 0 & 0 & 1 \end{bmatrix}$, and the vector of spectral coefficients would be $\begin{bmatrix} 1 & 1 & 0 & 0 \\ 0 & 1 & 0 & 0 \\ 0 & 0 & 1 & 1 \\ 0 & 0 & 0 & 1 \end{bmatrix} \begin{bmatrix} 0 \\ 1 \\ 1 \\ 0 \end{bmatrix} = \begin{bmatrix} 1 \\ 1 \\ 1 \\ 0 \end{bmatrix}$.

Note that transformation matrices are always invertible; this makes sense since polarity transformations can always be reversed. To emphasize the polarity, we can also denote $T$ as a polarity matrix $P_p$, where $p = 10$ in the former case, and $p = 20$ in the latter case; this notation will become useful in Section 3. Unfortunately, the vector of spectral coefficients is not in order from $p_0$ to $p_3$; therefore, we demonstrate in the next section how to map the spectral coefficients to their corresponding base functions for FPRM and KRM forms.

### C. Mapping FPRM and KRM Spectral Coefficients to Base Functions

We shall use the transformation introduced in the previous section as an example: $\begin{bmatrix} 1 & 1 & 0 & 0 \\ 0 & 1 & 0 & 0 \\ 1 & 1 & 1 & 1 \\ 0 & 1 & 0 & 1 \end{bmatrix} \begin{bmatrix} 0 \\ 1 \\ 1 \\ 0 \end{bmatrix} = \begin{bmatrix} 1 \\ 1 \\ 0 \\ 1 \end{bmatrix}$. Consider the polarity of the FPRM transformation: 10. For each spectral coefficient, compare the binary representation of that coefficient's index in the vector to the polarity; bits that match

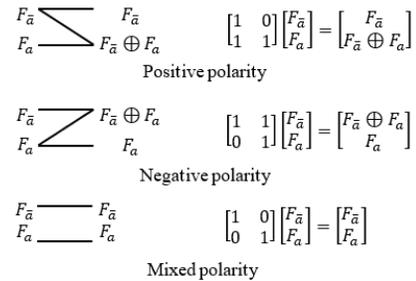

Fig. 2. *Butterfly kernels and their corresponding matrix polarity transformations*

represent variables that appear in that spectral coefficient's corresponding base function. Note that the spectral coefficients

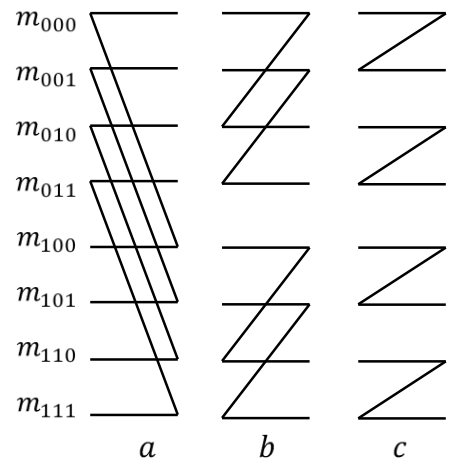

Fig. 3. *Example FPRM butterfly diagram for an arbitrary function $F(abc)$. This butterfly diagram generates the spectral coefficients for the FPRM form with polarity 100.*

can be considered to be arranged in an ordered matrix, where each coefficient has an index based on its position in the vector (the indices are 0-based). For example, for the first spectral coefficient (whose index has a binary representation of 00), we compare 00 to 10. Only the second bit matches, so the corresponding base function is $\bar{b}$. For the second spectral coefficient (whose index has a binary representation of 01), we compare 01 to 10. None of the bits match, so the corresponding base function is $1$. The next spectral coefficient has an index with a binary representation of 10; when compared to 10, both of the bits match, so the corresponding base function is $a\bar{b}$. Finally, the last spectral coefficient, with an index of 11, has the base function $a$. Ultimately, we find that the FPRM form is $F = (1)\bar{b} \oplus (1)1 \oplus (0)a\bar{b} \oplus (1)a$, which can be rearranged to $F = 1 \oplus \bar{b} \oplus a$. Our method is generalized to both FPRM and KRM forms below:

Method:

1. State the polarity of the Reed Muller form.
2. For each spectral coefficient, perform a digit-by-digit comparison of the index of the spectral coefficient to the polarity.
   a. Where digits match, the corresponding variable is part of the base function.
   b. If the polarity of a variable is 2 in the Reed Muller form (indicating mixed polarity), then the digit of the coefficient's index determines the polarity of the variable in the base function.

*D. FPRM and KRM Butterfly Diagrams and their relation to transformation matrices*

The polarity transformations from the previous section can also be represented with butterfly diagrams. Butterfly diagrams are evaluated from left to right; an EXOR is performed wherever two lines meet, and the value is propagated to the right otherwise. Positive polarity, negative polarity, and mixed polarity **butterfly kernels** are shown in Figure 2; there are multiple kernels in a butterfly diagram, organized into multiple columns.

A **set**, or **column**, of butterfly kernels is required for each variable. If positive Davio is performed on a variable, a set of positive butterfly kernels is used; if negative Davio is performed on a variable, a set of negative kernels is used; if Shannon is performed on a variable, a set of mixed kernels is used. For each column, its kernels are stretched by a power of two such that the kernels of each column are stretched by half the stretch of the previous column. For example, for the three-variable example butterfly with polarity 100 shown in Figure 3, the first column of butterflies for variable $a$ would contain 4 positive kernels, each stretched by 4 units; the next column for variable $b$ would contain 4 negative kernels, each stretched by 2 units; the final column for variable $c$ would contain another 4 negative kernels, each stretched by 1 unit (or, equivalently, "unstretched"). In this notation, butterflies are a representation of the polarity transformations previously discussed. Thus, each column of butterflies represents a transformation matrix, and the stretching of each set represents the Kronecker product. We use butterfly diagrams for two reasons: firstly, when implemented as part of an algorithm, their efficiency is much greater when compared to their corresponding matrix transformations; secondly, their graphical nature makes them more useful for visualizing FPRM/KRM transformations than their equivalent transformation matrices.

The classical FPRM minimization problem is to find the polarity (and the corresponding circuit) with minimum cost for a given Boolean function. We now provide an example of the generation of all FPRM polarities of a Boolean expression to demonstrate how FPRM aids in the optimization of Boolean expressions when they are realized as quantum circuits. Let us be given the expression $F = a + (b \oplus c)$, which has a minterm vector and Karnaugh map as shown in Figure 4. The original expression requires an OR gate, which is much more expensive than the EXOR gate when implemented in quantum.

$\begin{bmatrix} 0 \\ 1 \\ 1 \\ 0 \\ 1 \\ 1 \\ 1 \\ 1 \end{bmatrix}$

| ab \ c | 0 | 1 |
|---|---|---|
| **00** | 0 | 1 |
| **01** | 1 | 0 |
| **11** | 1 | 1 |
| **10** | 1 | 1 |

Fig. 4. *Minterm vector and Karnaugh map for the function $F(a, b, c) = a + (b \oplus c)$*

This is because the quantum implementation of the OR gate requires an Ancilla bit. In general, repeated usage of Ancilla bits is very expensive in quantum circuits. Thus, we seek to reduce the cost of the circuit by transforming the expression to an FPRM form, which does not require any OR gates.

As shown in Figure 5, we create all eight of the possible polarities of FPRM butterflies in order to generate all possible FPRM forms:

*E. Cost of FPRM/KRM Circuits with Various Polarities*

We continue to discuss the example as introduced in Figure 5. It can be determined which polarity is most optimized by calculating the cost of each polarity. The total cost can be determined by assigning costs to each component of a quantum circuit such that assigned costs are proportional to the real-life physical costs of building each component. For simplicity of the example, let us define the cost of a literal to be one; the cost of negation is considered negligible. We now provide an example of calculating total literal cost of an expression: for FPRM polarity 000, there are 5 literals — $\bar{a}$, $\bar{a}\bar{c}$ (2 literals), and $\bar{a}\bar{b}$ (2 literals). Therefore, the cost of the FPRM with polarity 000 would be 5. We then calculate the costs of all FPRM forms, which are summarized in Table 1. Polarities 001 and 010 are the cheapest, costing 4. We could choose any FPRM polarity and still have a less expensive implementation, but choosing polarities 001 or 010 would allow for the greatest reductions in cost.

Table 1. *Cost (in literals) of all FPRM polarities of* $F = a + (b \oplus c)$

| FPRM Polarity | Cost |
|---|---|
| 000 | 5 |
| 001 | 4 |
| 010 | 4 |
| 011 | 5 |
| 100 | 7 |
| 101 | 6 |
| 110 | 6 |
| 111 | 7 |

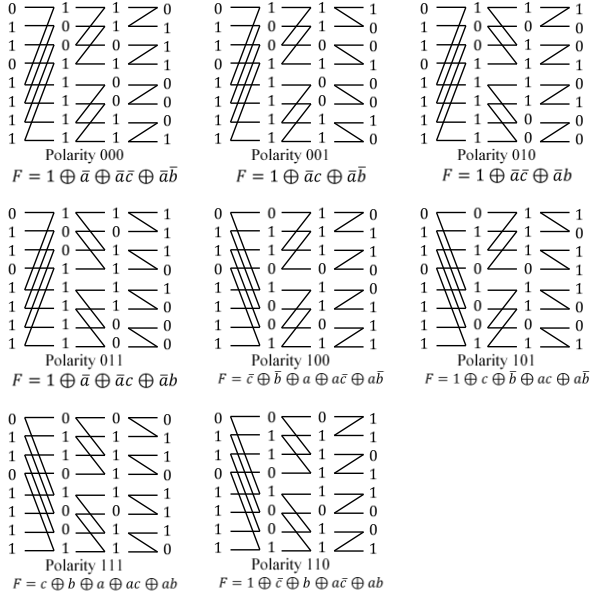

Fig. 5. *The eight possible FPRM butterfly diagrams for the function* $F(a,b,c) = a + (b \oplus c)$, *along with the eight FPRM polarity forms generated from the calculated spectral coefficients.*

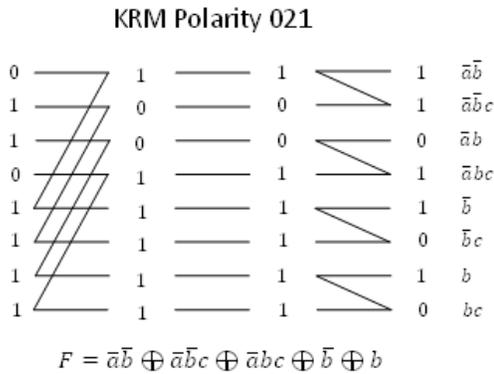

Fig. 6. *The KRM butterfly diagram for polarity 021, along with the KRM function generated from the butterfly, as determined by the spectral coefficient mapping method previously introduced.*

A similar idea is applicable to KRM forms. The cost KRM of polarities can be calculated in exactly the same way. Figure 6 shows the KRM polarity 021 of the same expression $F = a + (b \oplus c)$: counting literals, the cost of this polarity would be 10.

This example has demonstrated that FPRM and KRM polarities can be more optimized than the original function, and some polarities are further optimized than other polarities. We will now discuss quantum multiplexers before discussing how a similar application of butterfly diagrams can be used to optimize them.

*F. Quantum Multiplexers and Controlled Gates*

Let us first define the quantum multiplexer. A **quantum multiplexer** is a block of gates that will, based on a set of control variables (qubits), use a Boolean product of the control literals to select an arbitrary unitary quantum function (called a target function) that acts on a single target variable (qubit): see Figure 7. Note that in the diagram, the black circles denote "controls". That is, if the value on the line is $|1\rangle$ when it reaches the control, then the control is enabled. If all controls are enabled, then the associated target function will activate. Thus, when taken together, the controls of a target function form its **control function** (the previously mentioned product of control literals). For example, the function $F_0$ is controlled by both $\bar{c}_1$ and $\bar{c}_2$, which means that the control function is $\bar{c}_1\bar{c}_2$, and the target function is active if and only if $c_1 = 0$ and $c_2 = 0$. Target functions can be any single-qubit quantum functions. Thus, they are represented by unitary $2 \times 2$ matrices.

Suppose we have a multiplexer $M$ with $m$ control variables, which we can denote as $c_1, c_2, \ldots, c_m$. We can define any of the possible input states to the multiplexer in terms of the values on the control variables: the **input state** $i$ can be represented as a binary string, where each digit $i_k$ is equal to the value of its corresponding control variable $c_k$. For example, if we have $c_1, c_2, c_3$ which have values of 0, 1, 1 respectively, $i$ would be $011 = 3$. Note that there are only $2^m$ possible input states since we are assuming that the control variables are always in basis quantum states. However, we can be dealing with superposition states in the target qubit.

For the multiplexer $M$, let $F_i$ be the arbitrary single-qubit quantum function that will act on the target variable if and only if the input state is $i$: we can say that the function $F_i$ is a **target (controlled) function** that is **controlled by $i$**. For example, for the function $F_3 = F_{011}$ controlled by the three variables $c_1, c_2, c_3$, we say that $F_3$ is controlled by $i = 3 = 011$; this is equivalent to saying that $F_3$ is controlled by the Boolean product $\bar{c}_1 c_2 c_3$ since only the input state $i = 3$ satisfies this **control function**. It is clear that any multiplexer with $m$ control variables can be represented as $2^m$ controlled functions, one for each input state. Thus, we can uniquely represent the multiplexer as an ordered set:

$$M = \{F_0, F_1, \ldots, F_{2^m-1}\}_C$$

where $C$ is the ordered set of control variables. We choose to represent the multiplexer in this fashion for mathematical convenience in the "Proof" section below.

Note that a multiplexer of this form is a direct realization of a Minterm Sum of Products form: each minterm of the control variables controls its own function. For example, for a multiplexer with two control variables $c_1, c_2$, the minterms of the controls $\bar{c}_1 \bar{c}_2, \bar{c}_1 c_2, c_1 \bar{c}_2, c_1 c_2$ each appear once and each control a separate function $F_0, F_1, F_2, F_3$, respectively (once again, see Figure 7). Henceforth, this form of multiplexer that mirrors the Minterm Sum of Products form will be referred to as the **standard form**.

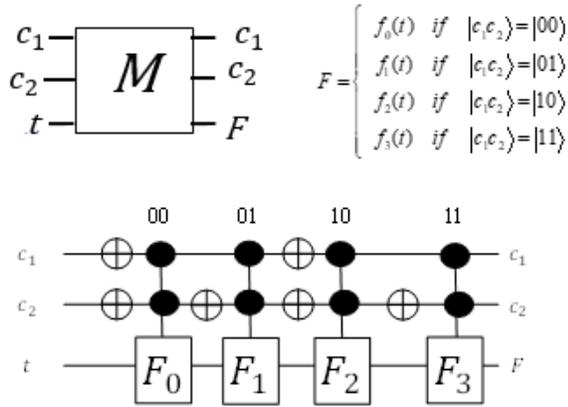

Fig. 7. *An arbitrary binary standard form quantum multiplexer with two controls, represented both mathematically and as a circuit diagram. The controls are listed above each function.*

### III. FIXED POLARITY QUANTUM FORMS (FPQF) AND KRONECKER QUANTUM FORMS (KQF)

#### A. Introduction to FPQF and KQF

We have previously discussed FPRM, a polarized form of a binary Boolean expression where each variable appears solely in either its complemented or uncomplemented form. We have also discussed KRM, a form of a binary Boolean expression where each variable appears in either positive, negative, and mixed polarities. For quantum multiplexers, given a set of control variables, $c_1, c_2, \ldots, c_m$, we can similarly introduce the concept of polarity. We can do so by creating a multiplexer where each control variable is in a fixed polarity; such a multiplexer will henceforth be referred to as a **Fixed Polarity Quantum Form (FPQF)**. Similarly, we can create a multiplexer where control variables are in either fixed or mixed polarity; such a multiplexer will be referred to as a **Kronecker Quantum Form (KQF)**. If we wish to create an FPQF or KQF multiplexer that realizes exactly the same functions as a particular standard form multiplexer for all input states while *fixing the polarities of some (or all) of the control variables*, it is clear that the functions $G_0, G_1, \ldots, G_{2^m-1}$ of the FPQF multiplexer will be different from the functions $F_0, F_1, \ldots, F_{2^m-1}$ of the standard form multiplexer since the control functions have changed: see Figure 8 for an example of the way that the control functions change. In an FPQF multiplexer, it is possible for multiple target functions to be active for a given input state; for example, for the input state $10$, both $G_0$ and $G_2$ are active. Figure 8 will also provide informal, analogical intuition for the similarities between FPRM forms and FPQF quantum multiplexer forms, and KRM forms and KQF quantum multiplexer forms: instead of spectral coefficients deciding whether or not a base function is expressed in FPRM/KRM, we have control variables deciding whether or not a target quantum function is expressed in FPQF/KQF.

Henceforth, we will denote target functions as $F_i$ if they are the target functions in a standard form multiplexer, and we will denote them as $G_i$ if they are the target functions of an FPQF/KQF multiplexer. First, it is useful to formally define how the target functions of a FPQF multiplexer $G_i$ operate.

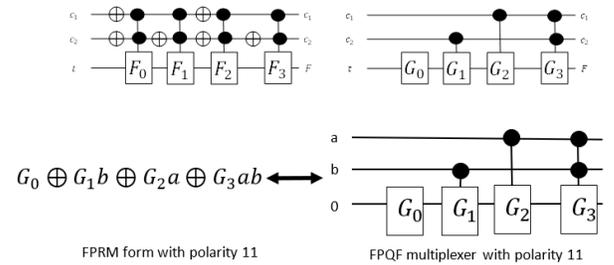

Fig. 8. *Comparison of standard form multiplexer and an FPQF multiplexer with polarity 11. This figure also demonstrates the analogical similarities between an FPRM form on a function of variables $a, b$ and an FPQF multiplexer with control variables $a, b$.*

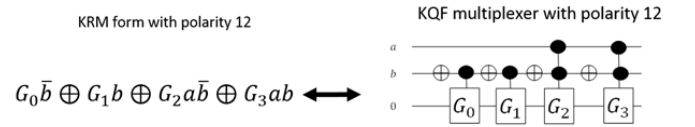

Fig. 9 *Comparison of the KRM form with polarity 12 to the KQF multiplexer with polarity 12. Note the similarities between the KRM and KQF forms.*

In an FPQF multiplexer, $G_i$ is a unitary matrix acting on the target variable and controlled by all input states $j$ such that $j_k = i_k$ if $i_k \neq 0$ (where $j_k$ and $i_k$ are the $k$th digits of $j$ and $i$, respectively). That is, $G_i$ is only controlled by the control variable $c_k$ if the $k$th digit of $i$ is not 0. We have previously mentioned that multiple target functions may be active for a particular input state; conversely, a target function can be activated by multiple input states. For example, on an example multiplexer with two controls, $G_1 = G_{01}$ is controlled by both the input states $01$ and $11$. That is, $G_1$ is activated if the input state is $01$ or $11$. Similar to our notation for a standard form multiplexer, we denote an FPQF multiplexer as $M = \{G_0, G_1, ..., G_{2^m-1}\}_{C,p}$ where $p$ is the polarity of the FPQF multiplexer as represented by a binary string. There are $2^m$ possible FPQF forms for a given set of functions $F_i$, so we need a fast way to compute all possible polarities of $[G_0, G_1, ..., G_{2^m-1}]$ from $[F_0, F_1, ..., F_{2^m-1}]$. Note that the circuit structures for all FPQF polarities are highly similar; while Figure 8 shows an FPQF multiplexer in polarity 11, other polarities can be realized simply by applying an inverter to the beginnings of control lines that we would like to place in negative polarity, then modifying the target functions $G_i$.

In a KQF multiplexer, whether or not a gate $G_i$ is active is dependent on the polarity. If the $k$th variable is in a fixed polarity, $G_i$ is controlled by all input states $j$ such that $j_k = i_k$ if $i_k \neq 0$. If the $k$th variable is in a mixed polarity, $G_i$ is controlled by all input states $j$ such that $j_k = i_k$. Negative fixed polarities can be realized simply by applying an inverter to the beginnings of controls lines that we would like to place in negative polarity, then modifying the target functions $G_i$. See Figure 9 for the comparison between a KRM form and KQF multiplexer.

There is a clear mathematical relationship between the target functions $F_i$ of the standard form multiplexer and the target functions $G_i$ of the FPQF multiplexer:

$F_0 = G_0$
$F_1 = G_1 \cdot G_0$
$F_2 = G_2 \cdot G_0$
$F_3 = G_3 \cdot G_2 \cdot G_1 \cdot G_0$

→

$G_0 = F_0$
$G_1 = F_1 \cdot F_0^{-1}$
$G_2 = F_2 \cdot F_0^{-1}$
$G_3 = F_3 \cdot F_1^{-1} \cdot F_0 \cdot F_2^{-1}$

Fig. 10. *Mathematical relationships between the target functions of a standard form multiplexer $F_i$ and the target functions of an FPQF multiplexer $G_i$ in polarity 11. This example case involves multiplexers with two control variables (and thus four target functions).*

Remember that a gate $F_i$ or $G_i$ is active if all controls are enabled (with value 1). Consider an example with two control variables, where we want the standard form and FPQF multiplexers to realize exactly the same functions. For input state 00, $F_0$ is active in the standard form multiplexer and $G_0$ is active in the FPQF multiplexer. Thus, $F_0 = G_0$. For input state 01, $F_1$ is active in the standard form multiplexer, but two gates are active in the FPQF multiplexer: $G_0, G_1$ (Figure 8 may help to visualize this). Thus, $F_1 = G_1 \cdot G_0$, and therefore $G_1 = F_1 \cdot G_0^{-1} = F_1 \cdot F_0^{-1}$. Note that the $\cdot$ operation is the matrix multiplication operation applied on the target functions when represented as unitary matrices. Similar observations can be made for input states 10 and 11; see Figure 10 for all the mathematical formulas that we derive. We have shown that the functions $G_i$ can be directly calculated from the functions $F_i$ for an example with two controls. In the next section, we demonstrate how to compose a transformation that can perform these calculations, which will enable us to generalize the process of calculating $G_i$ from $F_i$.

Similar observations can be made about the functions $G_i$ of KQF multiplexers. For a given multiplexer $M = \{F_0, F_1, ..., F_{2^m-1}\}_C$ and a given FPQF form $M = \{G_0, G_1, ..., G_{2^m-1}\}_{C,p}$, define $T_{m,p}$ as the transformation over $U$ that takes the vector $[F_0, F_1, ..., F_{2^m-1}]$ to the vector $[G_0, G_1, ..., G_{2^m-1}]$ for a certain polarity $p$ (remember that polarities are represented as binary strings in the same way as we previously introduced for FPRM polarities). We now provide a proof for decomposing $T_{m,p}$ into a Kronecker product of simpler transformations, which we can then represent with KQF butterfly diagrams similar to how we use KRM butterflies.

### B. *Proof for decomposition of FPQF polarity transformation to simpler transformations*

We first discuss a notation used in the following proof. Let us define an arbitrary transformation $B$, which acts on a set of two functions. We define the butterfly transformation $B_{n,i}$ as a transformation that acts on a set of $2^n$ functions represented as unitary matrices, $[F_0, F_1, ..., F_{2^n-1}]$, such that every set of functions $F_q$ and $F_r$, where the binary representations of $q$ and $r$ differ only at the $i$th digit, is acted upon by the transformation $B$. For example, if $B$ takes $[a, b]$ to $[b, a \cdot b]$, then $B_{2,1}$ is the transformation performed on $2^2 = 4$ functions, where $B$ is acted upon pairs of functions whose binary representation of index only differ at the 1st digit. The transformation $B_{2,1}$ would take the functions $[F_0, F_1, F_2, F_3]$ to $[F_1, F_0 \cdot F_1, F_3, F_2 \cdot F_3]$. This is because $B$ was applied to the pairs $[F_0, F_1] = [F_{00}, F_{01}]$ and $[F_2, F_3] = [F_{10}, F_{11}]$ since the indices of each pair of functions differ only at the 1st digit. Note that $B_{n,i}$ is the algebraic representation of a column of butterflies on $2^n$ objects where each butterfly kernel is stretched by $2^i$; thus, we call $B_{n,i}$ a **butterfly transformation**. We now show how a KQF transformation can be decomposed to three types of butterfly transformations: $P^1$, $P^0$, and $P^2$.

First, we define the transformation $P^1$, which takes $[a, b]$ to $[a, b \cdot a]$, the transformation $P^0$, which takes $[a, b]$ to $[b \cdot a, a]$, and the transformation $P^2$, which takes $[a, b]$ to $[a, b]$. From these transformations, define their respective butterfly transformations $P^1{}_{n,i}$, $P^0{}_{n,i}$, and $P^2{}_{n,i}$. We claim that $T_{m,p}{}^{-1} = P^{p_m}{}_{m,m} \cdot P^{p_{m-1}}{}_{m,m-1} \cdot ... \cdot P^{p_1}{}_{m,1}$, where $p^i$ is the

$ith$ digit of the targeted FPQF polarity $p$. We prove this by induction on $m$.

*1) Base Case: We first establish the claim with m = 1*

There are three possible polarities when $m = 1$. For polarity 0, we need to prove that $T_{1,0}^{-1} = P^0_{1,1}$, for polarity 1, we need to prove that $T_{1,1}^{-1} = P^1_{1,1}$, and for polarity 2, we need to prove that $T_{1,2}^{-1} = P^2_{1,1}$.

Polarity 1:

We have $M = \{F_0, F_1\}_C = \{G_0, G_1\}_{C,1}$. In order for this to hold, the two multiplexers must output the same values for inputs 0 and 1. We can verify that for input 0, we have $F_0 = G_0$ and that for input 1, we have $F_1 = G_1 \cdot G_0$.

Thus, $[F_0, F_1] = [G_0, G_1 \cdot G_0] = P^1_{1,1}([G_0, G_1])$. Since $T_{1,1}^{-1}$ is the transformation that takes $[G_0, G_1]$ to $[F_0, F_1]$, we have $T_{1,1}^{-1} = P^1_{1,1}$.

Polarity 0:

Here, we have $M = \{F_0, F_1\}_C = \{G_0, G_1\}_{C,0}$. For input 0, we have $F_0 = G_1 \cdot G_0$, and for input 1, we have $F_1 = G_0$.

Thus, $[F_0, F_1] = [G_1 \cdot G_0, G_0] = P^0_{1,1}([G_0, G_1])$ and $T_{1,0}^{-1} = P^0_{1,1}$.

Polarity 2:

Here, we have $M = \{F_0, F_1\}_C = \{G_0, G_1\}_{C,2}$. For input 0, we have $F_0 = G_0$, and for input 1, we have $F_1 = G_1$.

Thus, $[F_0, F_1] = [G_0, G_1] = P^2_{1,1}([G_0, G_1])$ and $T_{1,2}^{-1} = P^2_{1,1}$.

*2) **Induction**: We establish the claim for $m$, assuming that it holds for $m - 1$*

We divide our induction into two steps based on the leading digit, $p_m$, of the polarity.

Case 1: $p_m = 1$

First, note that for input states where the $mth$ control variable $c_m = 0$, the only controlled gates that can possibly be activated in the multiplexer $M = \{G_0, G_1, \ldots, G_{2^m-1}\}_{C,p}$ are $G_0, G_1, \ldots, G_{2^{m-1}-1}$. This is because none of the gates $G_0, G_1, \ldots, G_{2^{m-1}-1}$ are controlled by a control function that contains $c_m$. Thus, for input states $i$, $0 \le i \le 2^{m-1} - 1$, we can disregard the variable $c_m$. Let us denote the multiplexer that only involves input states $0 \le i \le 2^{m-1} - 1$ as $M'$.

If we let $M' = \{F_0, F_1, \ldots, F_{2^{m-1}-1}\}_{C'}$, $p' = p - 2^m$ ($p'$ is $p$ without its leading digit), and $C'$ as the set of control variables sans $c_m$, then $M' = \{G_0, G_1, \ldots, G_{2^{m-1}-1}\}_{C',p'}$. Thus, by definition, the transformation that takes $[G_0, G_1, \ldots, G_{2^{m-1}-1}]$ to $[F_0, F_1, \ldots, F_{2^{m-1}-1}]$ is $T_{m-1,p'}^{-1}$.

Now, consider the vector:

$$[G'_0, G'_1, \ldots, G'_{2^m-1}]$$
$$= T_{m-1,p'}([G_0, G_1, \ldots, G_{2^{m-1}-1}])$$
$$+ T_{m-1,p'}([G_{2^{m-1}}, G_{2^{m-1}+1}, \ldots, G_{2^m-1}])$$

Note that because $T_{m-1,p'}$ is composed of butterfly transformations, by the assumption of induction for $m - 1$:

$$[G'_0, G'_1, \ldots, G'_{2^m-1}]$$
$$= P^{p_{m-1}}_{m,m-1} \cdot P^{p_{m-2}}_{m,m-2} \cdot \ldots$$
$$\cdot P^{p_1}_{m,1}([G_0, G_1, \ldots, G_{2^m-1}])$$

Now, for all $i$, $0 \le i \le 2^{m-1} - 1$, we have shown that $F_i = G'_i$ by definition. Furthermore, for all inputs $i + 2^{m-1}$, $c_m = 1$ while all other control variables hold identical values, so each gate $G_{i+2^{m-1}}$ in multiplexer $M = \{G_0, G_1, \ldots, G_{2^m-1}\}_{C,p}$ is activated if and only if $G_i$ is activated. Therefore, $F_{i+2^{m-1}} = G'_{i+2^{m-1}} \cdot G'_i$, and $[F_i, F_{i+2^{m-1}}] = P^1([G'_i, G'_{i+2^{m-1}}])$. By definition, this is equivalent to

$$[F_0, F_1, \ldots, F_{2^m-1}] = P^1_{m,m}([G'_0, G'_1, \ldots, G'_{2^m-1}])$$
$$= P^1_{m,m} \cdot P^{p_{m-1}}_{m,m-1} \cdot \ldots$$
$$\cdot P^{p_1}_{m,1}([G_0, G_1, \ldots, G_{2^m-1}])$$

Thus, $T_{m,p}^{-1} = P^1_{m,m} \cdot P^{p_{m-1}}_{m,m-1} \cdot \ldots \cdot P^{p_1}_{m,1}$, completing the induction.

Case 2: $p_m = 0$

This case is essentially the same as case 1.

Case 3: $p_m = 2$

This case is essentially the same as case 1.

Now that we have proved $T_{m,p}^{-1} = P^{p_m}_{m,m} \cdot P^{p_{m-1}}_{m,m-1} \cdot \ldots \cdot P^{p_1}_{m,1}$, it is easy to determine $T_{m,p}$ by inverting our current equation: $T_{m,p} = A^{p_1}_{m,1} \cdot A^{p_2}_{m,2} \cdot \ldots \cdot A^{p_m}_{m,m}$, where $A^1 = (P^1)^{-1}$ takes $[a, b]$ to $[a, b \cdot a^{-1}]$, $A^0 = (P^0)^{-1}$ takes $[a, b]$ to $[b, a \cdot b^{-1}]$, and $A^2 = (P^2)^{-1}$ takes $[a, b]$ to $[a, b]$.

IV. DISCUSSION OF THE PROGRAM USED TO COMPUTE FPQFS

*A. FPQF and KQF Butterflies*

Here, we introduce the FPQF and KQF butterfly kernels that are used in FPQF and KQF butterflies. These are illustrated in Figure 11.

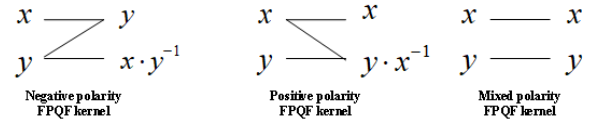

Fig. 11. *Butterfly kernels for FPQF and KQF butterfly diagrams. Only the first two types of kernels are present in FPQF diagrams.*

The construction of FPQF butterfly diagrams is similar to the construction of FPRM butterfly diagrams. However, note that the inputs to the diagram are no longer the minterms of a Boolean function, and the outputs are no longer spectral coefficients; rather, the inputs are the target functions $F_i$ of the standard form multiplexer, and the outputs are the target functions $G_i$ of the polarized multiplexer. See Figure 12 for an example of the FPQF butterfly diagrams for all polarities of FPQF on a multiplexer with three controls.

### B. Quantum Cost of FPQF/KQF Polarities

Having developed a technique to calculate all possible FPQF and KQF forms of a standard form multiplexer (that is, by constructing and evaluating all possible FPQF or KQF butterfly diagrams), we now introduce how the cost of quantum multiplexers will be calculated. Similar to [8], we define the cost of a quantum circuit in terms of the number of uncontrolled and single-controlled gates required to realize it; we can find this by summing the costs of all the controlled gates in the multiplexer circuit. Maslov et.al. [8] has previously established the following cost functions for Toffoli gates with $m$ controls (see Table 2). We will assume that we are able to use a single Ancilla bit for our entire circuit, so the equation $32m - 96$ is relevant to us. For reasons that are not important to the understanding of this paper (and which are discussed in [8], the construction of a multi-controlled $U$ gate (where $U$ is any arbitrary single-qubit quantum function) requires controlled- $\sqrt{U}$ gates and controlled- $\sqrt{U}^\dagger$ gates. It is theoretically possible to realize any $\sqrt{U}$ and $\sqrt{U}^\dagger$ gate; furthermore, it is also theoretically possible to realize singly-controlled versions of $\sqrt{U}$ and $\sqrt{U}^\dagger$. For these reasons, it is therefore theoretically possible to create any multi-controlled $U$ gate through the same process as the construction of a Toffoli (multi-controlled $NOT$) from singly-controlled and uncontrolled $\sqrt{NOT}$ and $\sqrt{NOT}^\dagger$ gates (Feynman gates are also required for all multi-controlled $U$ gates): see *Lemma 6.1* in [8]. Thus, the cost functions that have been developed by [8] for Toffoli gates can also be applied to formulate the cost of multi-controlled gates of any type.

### C. Pseudocode for the Program

A program was written to automate the process of generating FPQF butterflies and calculating costs of the resulting circuits. The program is written in Java because of its strong Object Oriented Programming (OOP) support; the code uses many classes (such as Multiplexer, Function, and more), and thus, Java's strong support of OOP is a large benefit. When executed, the program asks for the target functions of the standard form multiplexer in the form of unitary matrices as input (which are inputted in ascending natural order based off the controls), and outputs the least expensive polarity of FPQF. Alternatively, the program can also parse the names of common functions (e.g. "PX" for Pauli X, or "H" for Hadamard) to unitary matrices. The program uses a Depth First Search algorithm to generate all possible FPQF forms by recursively applying layers of butterflies; that is, for each layer of butterflies, the program first applies a layer of negative polarity butterflies to the multiplexer and recursively calls itself; after the call returns, the program backtracks by one layer and applies a layer of positive polarity butterflies to the multiplexer, then recursively calls itself. Butterflies are implemented as a series of matrix transformations, as seen in Figure 11. The negative polarity butterfly receives two unitary matrices, $x$ and $y$. It then outputs $y$ as the first output and $x \cdot y^{-1}$ as the second output; the latter output can be coded as the combination of a matrix inversion and a matrix multiplication. The positive polarity butterfly can be encoded similarly. Once all FPQF forms are generated, cost can be calculated as previously discussed: for each controlled function, determine the number of controls $n$ needed for that function. If $n < 10$, the program references the costs derived by [8]. Otherwise, the code uses the equation $32m - 96$ to determine the cost of the gate. By finding the costs of all FPQF and KQF polarities, the program can exhaustively determine which polarities are the cheapest, thus finding the exact minimum cost for an FPQF or KQF realization of a multiplexer.

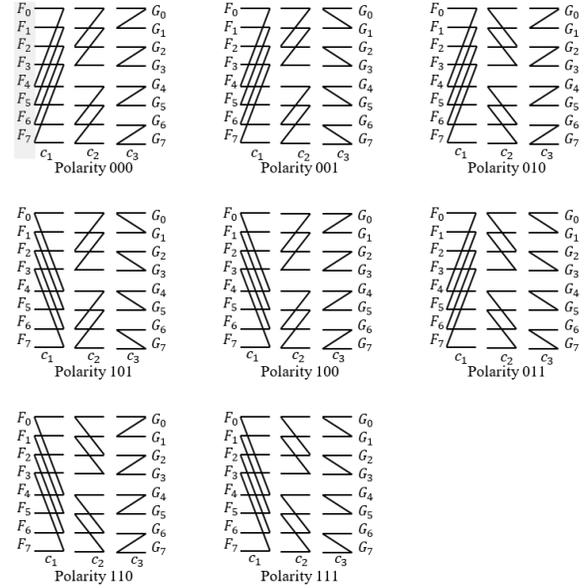

Fig. 12. *All polarities of FPQF butterflies for a standard form multiplexer with three control variables. Note that the sequences of functions $G_i$ generated by each FPQF butterfly are different.*

Table 2. *Costs of multi-controlled gates with $m$ controls. These gates use one or fewer Ancilla bits, which can be reused. Reproduced from [8].*

| Multi-controlled gate size $(m+1)$ | Number of Ancilla bits | Gate Cost |
|---|---|---|
| 1 | 0 | 1 |
| 2 | 0 | 1 |
| 3 | 0 | 5 |
| 4 | 0 | 13 |
| 5 | 0 | 29 |
| 6 | 1 | 52 |
| 7 | 1 | 84 |
| 8 | 1 | 116 |
| 9 | 1 | 154 |
| 10 | 1 | 192 |
| $(m+1) > 10$ | 1 | $32m - 96$ |

## V. ANALYSIS AND RESULTS

An additional program was written (also in Java) to generate large random test cases; initially, the test multiplexers were created by randomly pulling target functions from a set of common quantum functions: Pauli Rotations, Hadamard, NOT, V, V Hermitian, and Identity. A set of well-known quantum circuit cases were also created manually, such as a case that involves Identity, V, V, and NOT, as seen in Figure 13. The corresponding butterfly diagram for the case in Figure 13 can be found in Figure 14. On these well-known cases, the program correctly generated all polarities, including the optimal solutions.

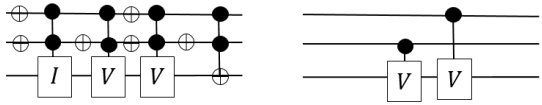

Fig. 13. *A standard form multiplexer (left) and its FPQF equivalent in polarity 11 (right). Note the drastic simplification of the circuit.*

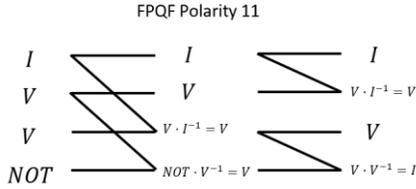

Fig. 14. *The butterfly diagram used to convert the standard form multiplexer to FPQF polarity 11 in Figure 13.*

### A. Randomly generated cases: Pauli X/Y/Z, Hadamard/Hadamard Hermitian, NOT, V/V Hermitian, Identity

When provided with larger, randomly generated cases, the program could run within 3 minutes or less on cases with up to 12 control variables. In the average case, the FPQF forms that were generated had costs as little as 30% of the cost of the standard form multiplexer, while the cost difference between the average polarity and the optimal polarity was often very small, less than 1% of the cost of the standard multiplexer: see Table 3 for results on large examples.

Average case cost reduction peaks near five or six control variables, then steadily decreases as the number of control variables (and thus, the complexity of the multiplexer) increases. This suggests that the effectiveness of the algorithm decreases as the complexity of the multiplexer increases, which is verified by data in Table 4.

Using KQFs did not reduce best case polarity cost at all; in test cases with 3-10 controls, the best polarity generated by Kronecker cost the same as the best polarity generated by FPQF forms. On the other hand, average case and worst case polarities became much more expensive. The KQF program was too slow to run on test cases with 11 or 12 controls. These results seem to indicate that on randomly generated data, KQFs are not more effective for optimization than FPQF forms.

Table 4. *Data on very large cases generated with Paulis, Hadamards, NOTs, V/V Hermitians, and Identities. Because of runtime constraints, the program generated a random polarity for these cases with more controls.*

| Number of Controls | Original Cost | Random Polarity Cost | Cost Reduction |
|---|---|---|---|
| 13 | 220840 | 82519 | 63% |
| 14 | 4825863 | 1988962 | 59% |
| 15 | 10501260 | 4520740 | 57% |
| 16 | 22759026 | 10139597 | 55% |
| 17 | 49036637 | 22488539 | 54% |
| 17 | 49079292 | 22479991 | 54% |

Table 3. *Data on cases generated with Paulis, Hadamards, NOTs, V/V Hermitians, and Identities. For these cases with fewer controls, the program exhaustively generated all polarities, finding the best case, worst case, and average case costs.*

| Number of Controls | Original Cost | Best Polarity | Worst Polarity | Average Polarity | Average Cost Reduction |
|---|---|---|---|---|---|
| 3 | 98 | 29 | 38 | 33 | 66% |
| 4 | 390 | 93 | 131 | 120 | 69% |
| 5 | 1431 | 312 | 414 | 380 | 73% |
| 6 | 4505 | 1169 | 1236 | 1215 | 73% |
| 7 | 12519 | 3326 | 3506 | 3444 | 72% |
| 8 | 32240 | 9226 | 9508 | 9414 | 71% |
| 9 | 82218 | 24207 | 24809 | 24589 | 70% |
| 9 | 80481 | 24237 | 24807 | 24613 | 69% |
| 10 | 193725 | 61203 | 62601 | 62168 | 68% |
| 10 | 192375 | 61588 | 62611 | 62170 | 68% |
| 11 | 450778 | 151438 | 153630 | 152807 | 66% |
| 11 | 432274 | 151801 | 153699 | 152866 | 65% |
| 11 | 438699 | 151739 | 153673 | 152846 | 65% |
| 12 | 977109 | 364440 | 368021 | 366688 | 62% |
| 12 | 986646 | 364788 | 368111 | 366718 | 63% |

Note that in many of these test cases, the cost of both the average case polarity and the worst case polarity of the FPQF forms are very close to the cost of the best polarity, and thus, the FPQF algorithm could be rewritten to work on much larger sizes of multiplexer if, instead of searching for the most optimal FPQF polarity, the program computes a random polarity, relying on the average case performance to provide a good multiplexer. However, the larger test cases are all randomly generated, which means that there may be large, structured usages of multiplexers where the optimal polarity costs significantly less than the average polarity; thus, randomly choosing a polarity will work well on completely random cases, but can potentially have poor results in large, structured cases.

We modified our algorithm to calculate a single, random polarity so that we would be able to test larger cases (otherwise, runtimes could exceed several hours, if not days). It can be seen that while the cost reduction is still significant, it decreases as the number of control variables increases. Given that testing on smaller cases has suggested that random polarities (the average case) are very close to the cost of the best case polarity, this data suggests that the effectiveness of the FPQF method decreases as the size of the multiplexer increases. Since the KQF method has been shown to have significantly weaker average case polarity costs, it is not meaningful to test random polarity generation on larger cases.

Table 5. *Data on cases generated with NOTs and V/V Hermitians. Once more, the program generated all polarities to find the best, worst, and average case costs.*

| Number of Controls | Original Cost | Best Polarity | Worst Polarity | Average Polarity | Average Cost Reduction |
|---|---|---|---|---|---|
| 5 | 1696 | 306 | 400 | 348 | 79% |
| 6 | 5440 | 877 | 1086 | 982 | 82% |
| 6 | 5440 | 890 | 1172 | 1016 | 81% |
| 7 | 14976 | 2230 | 2820 | 2579 | 83% |
| 8 | 39680 | 6346 | 8293 | 7191 | 82% |
| 9 | 98816 | 16802 | 20691 | 18909 | 81% |
| 10 | 230400 | 43386 | 50301 | 47182 | 80% |
| 11 | 526336 | 109087 | 120380 | 115236 | 78% |
| 12 | 1183744 | 268755 | 287182 | 277466 | 77% |
| 12 | 1183744 | 266766 | 286449 | 277383 | 77% |
| 12 | 1183744 | 266781 | 287745 | 277013 | 77% |

Table 6. *Data on cases generated with NOTs and V/V Hermitians. Once more, the program generated a random polarity because of runtime constraints.*

| Number of Controls | Original Cost | Random Polarity Cost | Cost Reduction |
|---|---|---|---|
| 13 | 2629632 | 651038 | 75% |
| 14 | 5783552 | 1508962 | 74% |
| 15 | 12615680 | 3426441 | 73% |
| 15 | 12615680 | 3413084 | 73% |
| 16 | 27328512 | 7643100 | 72% |
| 16 | 27328512 | 7661476 | 72% |
| 17 | 58851328 | 16981531 | 71% |
| 17 | 58851328 | 16952012 | 71% |

### B. Randomly generated cases: NOT, V/V Hermitian

This method was also tested on quantum multiplexers randomly generated from a smaller pool of target gates: NOT, V, and V Hermitian only. We are interested in using this set of gates as targets because there are several algorithms to synthesize quantum reversible circuits with this set of gates [8]. It can be seen in Table 5 that the cost difference between the best and average case polarities is slightly greater; in these cases with a smaller pool of target gates, it may be more useful to search for the best polarity instead of only using a random one. Observe once more that randomly generated functions are the most difficult cases; therefore, if the intent of the designer was to use many negative controls (as seen in Grover algorithm, for example), a randomly chosen polarity may be significantly worse than the minimum solution.

Once more, the KQF method did not offer improvements on the best case cost. In the 5 control case (the first row of Table 5), the best polarity case found by the KQF method has cost 297, versus FPQF's cost of 306; this is the only case where KQF offered an improvement. Similar to the set of data in *A*, KQF's average and worst case costs were significantly worse than those of FPQF, and the runtime was much longer.

Like previously, we modified the program to find random polarities for cases with greater numbers of controls because of runtime constraints. Again, note the trend of decreasing effectiveness of the algorithm as seen in Table 6. Once again, it is not meaningful to test random generation with the KQF method on larger cases.

### C. PLA Benchmarks: FPRM test cases (NOT, Identity)

Additional testing was run on PLA benchmarks; these industrial benchmarks are taken from classical logic synthesis of Boolean functions. They were converted to quantum multiplexers by converting the Karnaugh maps to standard form multiplexers (a Karnaugh map can easily be represented as a Minterm ESOP, which can be easily converted into a standard form multiplexer). On this benchmark data, there were marked differences in performance between FPQF and KQF. As previously noted, KQF runs much slower than FPQF, but this time it generated cheaper multiplexers in many cases. Notable cases include Sao2f1, Sao2f2, and Sao2f3, where KQF was significantly cheaper than FPQF. The data suggest that KQF has an effective use on industry circuits since benchmarks correspond to these types of real life circuits.

Note that these types of test cases are much more reflective of KQF's performance on real-life data than the randomly generated cases introduced and discussed in A. and B. This is because real-world applications of KQF often involve highly organized data. Real-life circuits will more frequently use multiplexers that cover disjoint minterms with high Hamming distance. For example, consider the multiplexer that realizes the ESOP function $F = abc \oplus \bar{a}\bar{b}\bar{c}$. The FPQF method would create a highly expensive multiplexer (more expensive than the original standard form multiplexer), while the KQF method would simply find the best solution: the original multiplexer. However, KQF still remains useful towards optimization. If there were a 4-variable function such as $F = \bar{a}\bar{b}\bar{c}\bar{d} \oplus \bar{a}\bar{b}\bar{c}d \oplus abc\bar{d} \oplus abcd$, the cost of the original multiplexer is much higher than the cost of the KQF form: $F = abc \oplus \bar{a}\bar{b}\bar{c}$.

## VI. CONCLUSION

In this paper, we first define standard form multiplexers and two new types of multiplexer forms: FPQF and KQF forms. Next, we provide a method to convert a standard form multiplexer to an FPQF or KQF form using a polarity transformation, and we give a proof for decomposing these transformations into smaller transformations that can be represented with butterfly diagrams in an analogous way to well-known FPRM and KRM butterfly diagrams. Then, we test this method for randomly generated standard form multiplexers that use target functions from a small pool of common quantum functions, finding that cost reductions are as high as 80%, but steadily decrease as size (and equivalently, complexity) of the standard multiplexer increases. We also find that while it was originally thought that one should use Depth First Search to find the maximally optimized FPQF form for each standard form multiplexer, data on smaller randomly generated multiplexers suggests that the small cost difference between the best case polarity and average case polarity do not justify the extremely long runtime needed to find the maximally optimized FPQF form; instead, finding a random polarity of FPQF form yields nearly identical cost reductions when given highly complex random circuits. This, however, does not apply to non-random circuits that may result from previous stages of quantum compilation. Additionally, we discovered that on randomly generated data, KQF offers no benefit on best case cost reduction, while having a longer run time and a lesser average case cost reduction. This suggests that on randomly generated multiplexers, there is no benefit to using KQF; only FPQF should be used to optimize these types of multiplexers.

Table 7. *Data on PLA benchmarks. These were converted from classical logic synthesis benchmarks.*

| Benchmark Name | Original Cost | FPQF Best Polarity | KQF Best Polarity |
|---|---|---|---|
| 9sym_d | 81060 | 4340 | 4340 |
| Con1f1 | 7956 | 165 | 155 |
| Con2f2 | 10296 | 85 | 74 |
| Exam1_d | 56 | 6 | 6 |
| Exam3_d | 180 | 24 | 24 |
| Life_d | 27020 | 178 | 176 |
| Max46_d | 11966 | 10927 | 10927 |
| Newill_d | 22010 | 936 | 935 |
| Newtag_d | 36270 | 364 | 353 |
| Rd53f1 | 318 | 150 | 150 |
| Rd53f2 | 1060 | 60 | 60 |
| Rd53f3 | 848 | 10 | 10 |
| Rd73f1 | 7488 | 126 | 126 |
| Rd73f2 | 7488 | 14 | 14 |
| Rd73f3 | 7488 | 1050 | 1050 |
| Rd84f1 | 18600 | 168 | 168 |
| Rd84f2 | 19840 | 16 | 16 |
| Rd84f3 | 155 | 155 | 155 |
| Rd84f4 | 25110 | 2100 | 2100 |
| Sao2f1 | 4050 | 4144 | 3501 |
| Sao2f2 | 4500 | 6450 | 4422 |
| Sao2f3 | 107100 | 5579 | 4504 |
| Sao2f4 | 52425 | 6352 | 5746 |
| Sym10_d | 188325 | 15430 | 15430 |
| Xor5_d | 848 | 10 | 10 |

Table 8. *Summary of the conclusions of the paper.*

| Type of Data | Comparison being made | Conclusions/Results |
|---|---|---|
| Optimization process on all types of benchmark data (as seen below) | Original (no optimization) vs FPQF/KQF (optimization) | FPQF/KQF multiplexers were almost always better than the original. However, for two PLA benchmarks, FPQF had a worse cost than original. KQF was always better or equal in terms of multiplexer cost than the original. |
| Randomly generated data that uses the gates $P_X P_Y P_Z$ HH+ VV+ NOT I | FPQF vs KQF | FPQF and KQF always had the same best case cost, but FPQF always had better average case costs, and the FPQF program also ran significantly quicker. Thus, the FPQF program always performed better than the KQF program. |
| Randomly generated data that uses the gates VV+ NOT | FPQF vs KQF | FPQF and KQF almost always had the same best case cost; in one case, KQF netted a tiny, 1% reduction in best polarity cost. However, FPQF always had better average case costs, and the FPQF program also ran significantly quicker. Thus, the FPQF program always performed better than the KQF program with one tiny exception. |
| PLA FPRM Benchmarks | FPQF vs KQF | The KQF program ran slower. However, in roughly half of the cases, KQF generated a cheaper best case polarity than FPQF. Thus, it is meaningful to use the KQF program instead of the FPQF program on this type of real-world data. |
| In general (based on all data | FPQF vs KQF | It is meaningful to use the KQF program instead of the FPQF program since it **never** generates worse results than FPQF. However, sometimes it may generate better results. The only reason not to use KQF is on extremely large multiplexers where KQF could not run in a practical amount of time. |

Additionally, we tested our FPQF and KQF methods on PLA industry benchmarks (FPRM benchmarks that we converted into standard form multiplexers). We discovered that for several of these benchmarks, the KQF method results in a cheaper, sometimes significantly cheaper, best case multiplexer than the FPQF method. This suggests that for real-world data, it may be worthwhile to use the KQF method over the FPQF method since it does a more complete search than FPQF (albeit at the expense of a longer runtime).

The FPQF and KQF methods suggest that conversion of a standard form multiplexer to an FPQF or KQF form can also be extended to multivalued quantum multiplexers: it is expected that the effectiveness and cost reductions of these methods will be very high for multivalued logic as well. Furthermore, the steadily decreasing trend of effectiveness of this method on larger randomly generated multiplexers suggests that more research should be conducted to find alternate methods for optimizing larger standard form multiplexers since it appears that the FPQF method will not net many savings at extremely high sizes. This is consistent with findings in [11] and [12], which suggest that Reed Muller-based methods are highly ineffective at minimizing functions that contain minterms with high Hamming distance.

A summary of our conclusions can be found in Table 8.

## VII. REFERENCES

[1] R. R. Tucci, "A Rudimentary Quantum Compiler (2cnd Ed.)," arXiv, 2008.

[2] J. J. Vartiainen, M. Möttönen and M. M. Salomaa, "Efficient decomposition of quantum gates," arXiv, 2008.

[3] V. V. Shende, S. S. Bullock and I. L. Markov, "Synthesis of Quantum Logic Circuits," arXiv, 2006.

[4] V. V. Shende, I. L. Markov and S. S. Bullock, "Minimal Universal Two-Qubit CNOT-based Circuits," arXiv, 2004.

[5] M. Davio, J.-P. Deschamps and A. Thayse, Discrete and Switching Functions, St-Saphorin: Georgi Publishing Company, 1978.

[6] M. Perkowski and D. Shah, Design of Regular Reversible Quantum Circuits, Portland: Portland State University, 2010.

[7] D. Maslov and G. W. Dueck, "Improved Quantum Cost for n-bit Toffoli Gates," arXiv, 2004.

[8] A. Barenco, C. H. Bennett, R. Cleve, D. P. DiVincenzo, N. Margolus, P. Shor, T. Sleator, J. Smolin and H. Weinfurter, "Elementary gates for quantum computation," arXiv, 1995.

[9] S. R. Hutsell, "An Eigenanalysis and Synthesis of Unitary Operators used in Quantum Computing Algorithms," Portland State University, PHD dissertation, Portland, 2009.

[10] D. H. Green, "Families of Reed-Muller canonical forms," *International Journal of Electronics,* vol. 70, no. 2, pp. 259-280, 1991.

[11] A. Mishchenko and M. Perkowski, "Fast Heuristic Minimization of Exclusive-Sums-of-Products," *5th International Reed-Muller Workshop,* 2001.

[12] L. Csanky, M. A. Perkowski and I. Schafer, "Canonical restricted mixed-polarity exclusive-OR sums of products and the efficient algorithm for their minimisation," *IEE Proceedings E - Computers and Digital Techniques,* vol. 140, no. 1, pp. 69-77, 1993.